\documentclass[review]{elsarticle}
\usepackage{hyperref}
\usepackage{amsmath}
\usepackage{amssymb}
\usepackage{graphicx}
\usepackage{subfig}
\usepackage{geometry}
\usepackage{float}
\usepackage{nomencl}
\journal{Journal of \LaTeX\ Templates}
\begin{document}
\begin{frontmatter}
\title{Computational Exploration of Inclined Magnetic Fields and Variable Thermal Flux Effects on the Flow of Dusty Hybrid Nanofluid around Stretching/Shrinking Wedge}
\author[label1]{N. Muqaddass \corref{mycorrespondingauthor}}\ead{nimra.muqaddass@imtlucca.it}
\address[label1]{MUSAM - Multi\_scale Analysis of Materials, IMT School for Advanced Studies Lucca, Italy}
\author[label2]{W. A. Khan}\ead{wkhan1956@gmail.com}
\address[label2]{Department of Mechanical Engineering, College of Engineering, Prince Mohammad Bin Fahd University, Al Khobar 31952, Kingdom of Saudi Arabia}
\author[label3]{F. Mabood}\ead{fmabood@fanshawec.ca}
\address[label3]{School of Information Technology, Fanshawe College London, ON, Canada}
\cortext[mycorrespondingauthor]{Corresponding author}
\begin{abstract}
This extensive investigation explores the influence of inclined magnetic fields and radiative non-linear heat flux on the behavior of dusty hybrid nanofluids over stretching/shrinking wedges. Employing $Cu$-$SiO_2$ as a hybrid nanoparticle composition and ethylene glycol $(EG)$ as the base liquid, the study investigates the fluid's response to a uniform magnetic field. The governing partial differential equations and associated boundary conditions are adeptly transformed into ordinary differential equations using appropriate transformations and then non-dimensionalized. Numerical simulations are executed using MATLAB and the bvp-4c solver. The outcomes offer a profound insight into thermofluid dynamics in industrial applications featuring intricate fluid flows, evaluating the influence of magnetic parameters on diverse fluid types, including nanofluids and dusty hybrid nanofluids. Furthermore, the investigation analyzes the impact of heat production and absorption on both vertical and horizontal plates, studying the significance of the velocity ratio factor in relation to the drag coefficient and local Nusselt number under thermal conditions of generation and absorption.
\end{abstract}
\end{frontmatter}
\section{Introduction}

\begin{figure}[h]
\centering
\includegraphics[width=11cm, height=9cm]{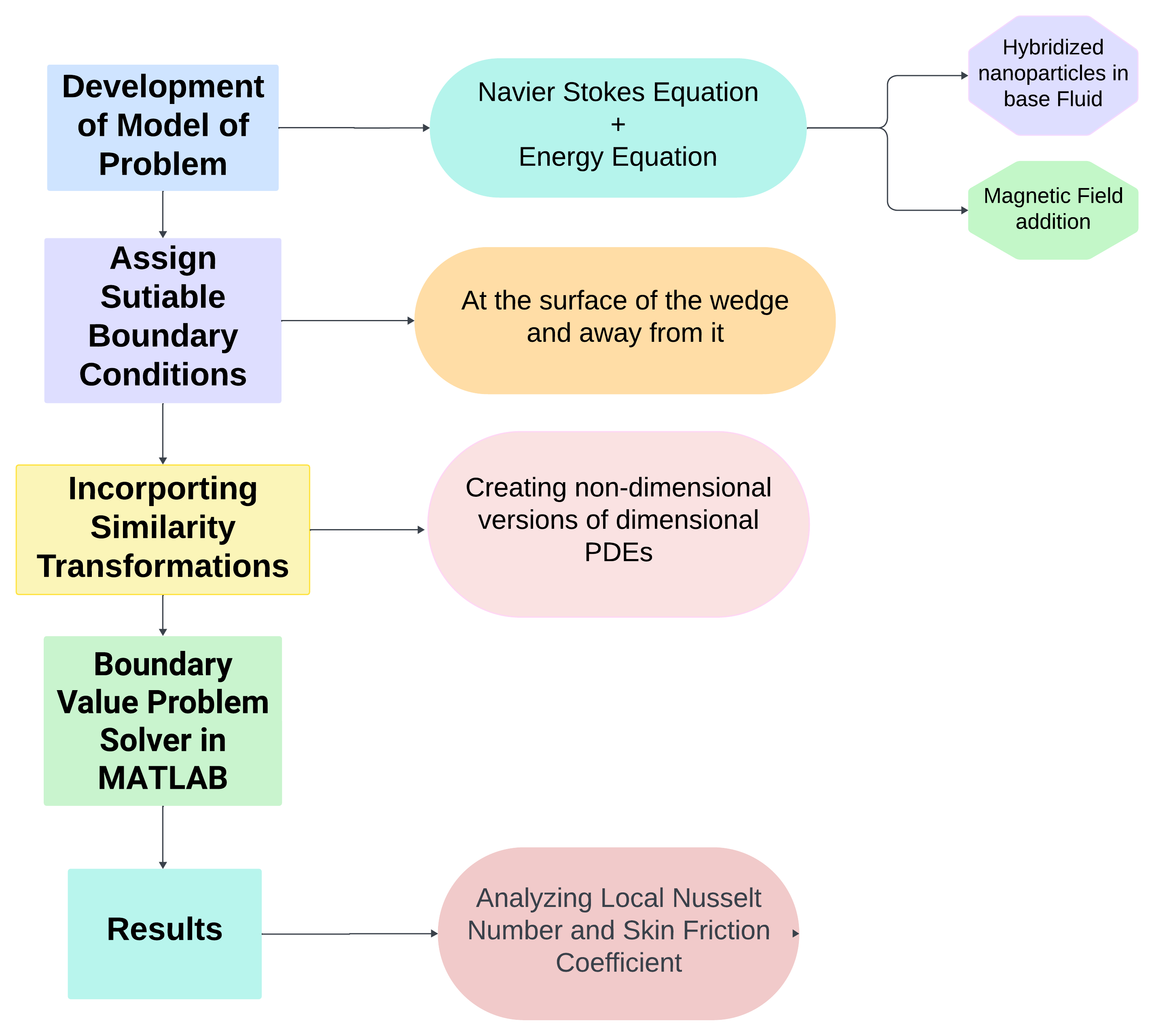}
\caption{An overview of the steps involved in numerical analysis.}
\label{fig1}
\end{figure}

In recent years, the exploration of advanced nanofluid formulations has captivated considerable interest due to their potential to enhance thermal and rheological properties across diverse engineering applications. Among these formulations, the integration of different nanoparticle types into a single fluid, known as hybrid nanofluids, has emerged as a promising avenue for achieving tailored and improved functionalities. Various base fluids are frequently used in nanofluids, including water, ethylene, triethylene glycol, refrigerants, lubricants, oils, polymer blends, and Biological fluids. Alumina, copper, zirconia, gold, silver, titania, $CuO$, $SiO_2$, $Al_2O_3$, metal carbides, metal nitrides, $SiN$, crystalline carbon, diamond, carbon nanotubes, and other chemically adapted materials play a significant role. Among their distinct features, nanofluids display uniformity, exceptional thermal transport at reduced particle clustering, stability over extended periods, and minimum clogging. These properties make them particularly interesting for a variety of applications. Choi's seminal work \cite{intro1} paved the way for extensive research into nanofluids, and the concept of leveraging nanoparticles to enhance fluid properties, particularly thermal conductivity, has since become a widely explored area by numerous researchers \cite{intro2, intro3, intro4, intro5}. The study conducted by \cite{intro5b} analyzed the effects of non-linear radioactive heating on copper and silver nanoparticles when combined with water-based convection. \cite{intro5a} found that the presence of magnetization leads to a higher energy profile but a decrease in velocity in an aqueous nanofluid containing $Cu$ and $Al_{2}O_{3}$. \cite{intro5c} examined the behavior of two different nanofluids containing graphene oxide nanoparticles and a mixture of $H_2O$ and $(Ch_2OH)_2$ as they flowed over a stretching cylinder, finding that increasing the dynamic viscosity of the nanofluids resulted in a decrease in temperature distribution. \cite{intro5e} used the Buongiorno model to investigate the behavior of a Maxwell nanoliquid as it moved between two horizontally oriented, stretchable, rotating disks. \cite{intro5d} studied the incompressible, stable power law nanofluid gyrotactic microorganisms flow, which involves the energy-carrying gyrotactic microbe that moves between parallel plates. \cite{intro5f} discovered a dual impact on axial velocity for different shapes of copper and copper oxide nanoparticles using water as a base fluid between two parallel plates. By taking into account the effects of Brownian motion, thermophoresis, and non-linear heat radiation, \cite{intro5g} performed a numerical study of nanofluid flow over a heated spinning disc with the use of Arrhenius activation energy. Recent research has yielded several articles concerning the properties of nanofluids as they respond to varying physical conditions \cite{intro5x, intro5y, intro5z, intro5z1, intro5z2, intro5z3}.
While nanofluids exhibit the potential to enhance thermal efficiency, researchers are persistently exploring more effective alternatives. Fortunately, recent strides in science and technology have given rise to the development of hybrid nanofluids, incorporating two different forms of nanoparticles into the base fluid. The innovative formulation presented here seeks to harness the advantages of multiple nanoparticle types or sizes within a single fluid, aiming to surpass the performance of conventional single-component nanofluids. \cite{intro8} delved into the flow of an ethylene glycol-H2O solution suspended with hybrid nanoparticles and found that the heat flux is contingent on the geometric coefficient. \cite{intro7} scrutinized the characteristics of the thermal transmission process by considering three distinct nanomolecules—alumina, titania, and copper on the flowing of the Cattaneo–Christov thermal flow pattern throughout an infinitely stretching plate. \cite{intro6} explored the significance of a horizontal electromagnetic field on the flow of an $Al_2O_3$-water nanofluid through three parallel fins surrounding a moderately heated hexagonal chamber. \cite{intro9a} conducted a study to enhance energy and mass transfer rates in manufacturing and technological applications, utilizing a ternary hybrid nanofluid consisting of titanium dioxide $(TiO_{2})$, aluminum oxide $(Al_{2}O_{3})$, and silicon dioxide $(SiO_{2})$ in a glycol/water carrier fluid. The study employed the squeezing of parallel infinite plates. \cite{intro9} observed that when a hybrid ferro-fluid is subjected to mass suction and partial slip, reduced shear stress and high heat transfer rates are achieved. This occurs when Stefan suction/blowing impacts and a temporal magnetism in the crosswise flow direction constantly influence the fluid. Several researchers have recently reported significant advancements in the flow of ternary hybrid nanofluids \cite{intro2023a, intro2023b, intro2023c, intro2023d, intro2024, intro2020a, intro2020b, intro2022a, intro2012c}.
This paper extends the research conducted by \cite{7}, which focused on analyzing the heat transmission features of a water-based dusty hybrid nanofluid comprised of $Fe_{2}SO_{4}$ and $Cu$ over a horizontal stretching sheet with the influence of heat generation and absorption. The current study breaks new ground by examining a hybridized dusty nanofluid $Cu$-$SiO_{2}$-$EG$ around a stretching/shrinking wedge body subjected to an inclined magnetic field and nonlinear heat flux. This investigation introduces a steady problem and employs a convergent numerical technique (as indicated in Figure \ref{fig1}) to guarantee the precision and reliability of the results. The primary objective is to enhance the understanding of the heat transfer mechanism in dusty hybrid nanofluids, offering valuable insights for engineers and researchers. It is emphasized that this investigation is an original contribution, and the presented numerical computations have not been published previously. Given the widespread application of dusty hybrid nanofluids formed of ethylene glycol ($EG$) as the base liquid and $Cu$-$SiO_{2}$ nanoparticles in various industrial and engineering settings, the findings are anticipated to be highly valuable.
\section{Problem Formulation}
\begin{figure}[h]
\centering
\includegraphics[width=14cm, height=9cm]{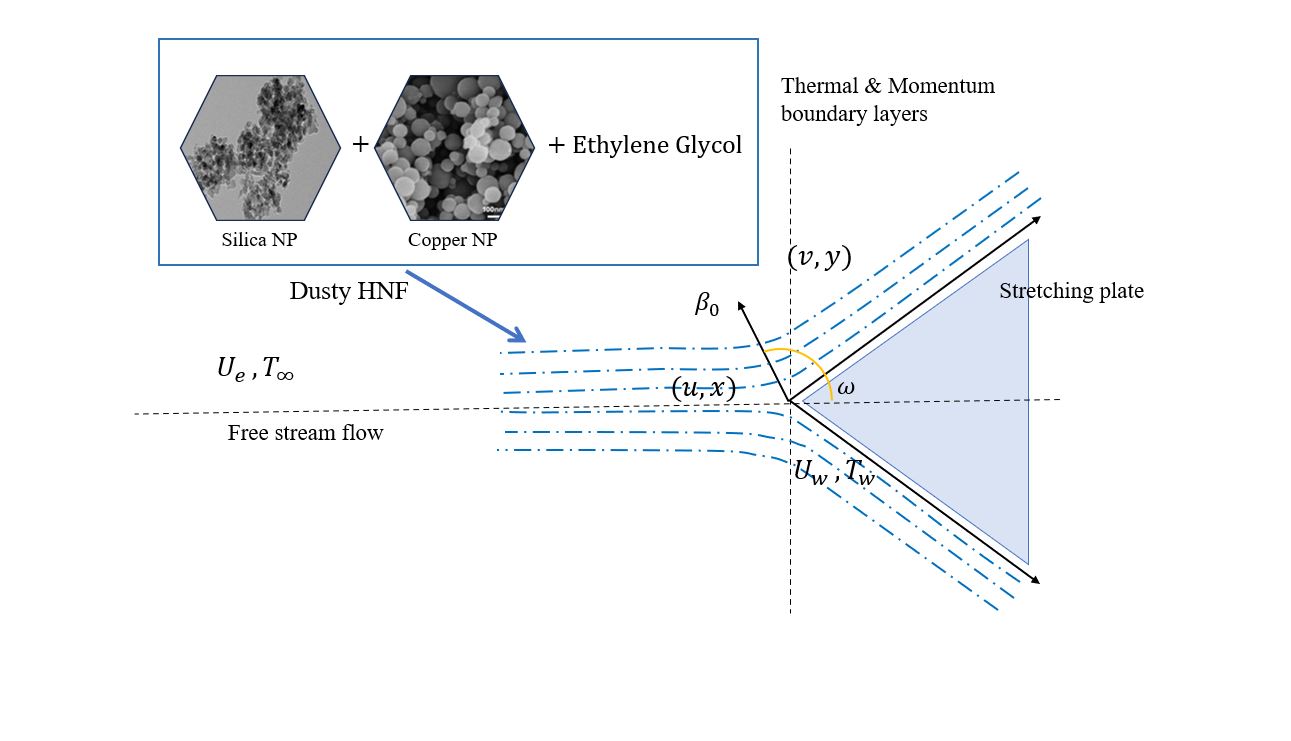}
\caption{The Schematic representation of the problem.}
\label{figdiagram}
\end{figure}

A two-phase flow of incompressible hybrid nanofluid, comprising Copper and Silicon dioxide nanoparticles in Ethylene glycol, is considered over a stretching/shrinking wedge surface. A uniform magnetic field $(\beta_{0})$ is applied over the wedge at $y>0$, inclined at an angle $\omega$. The free stream velocity and velocity of surface movement are denoted as $u_{e}(x)=U_{e}x^{m}$ and $u_{w}(x)=U_{w}x^{m}$, correspondingly. Here, $U_{w}>0$, $U_{w}<0$, and $U_{w}=0$ correspond to the stretching, shrinking, and static case, respectively, with $U_{e}$ being a positive constant (see Figure \ref{figdiagram}). The wedge angle is defined by $m=\beta /(2-\beta) \in [0,1]$, where $\beta$ represents the Hartree pressure gradient, and $\Omega=\beta \pi$ signifies the overall angular span. Additionally, $m=0$ ($\beta=0$) and $m=1$ ($\beta=1$) correspond to the boundary layer above a flat flat surface $(\Omega=0)$ and the stagnation point flow on a vertical flat plate $(\Omega=\pi)$, respectively. The system of equations governing the momentum and energy characteristics for the fluid and dust particle phases is established \cite{1, 2, 3, 4, 5}.

For fluid phase:
\begin{equation}
\label{dim_continuity}
\frac{\partial {u}}{\partial {x}}=-\frac{\partial {v}}{\partial {y}},
\end{equation}
\begin{equation}
\label{dim-v}
\rho_{hnf} \bigg( {u} \frac{\partial {u}}{\partial {x}}+{v} \frac{\partial {u}}{\partial {y}} \bigg)=\mu_{hnf} \frac{\partial^{2}{u}}{\partial {y}^{2}} -  \delta_{hnf} \beta_{0}^{2} Sin^{2} \omega \big(u-u_{e}(x) \big) + KN(u_{p}-u),
\end{equation}
\begin{equation}
\label{dim-temp}
(\rho C_p)_{hnf} \bigg({u}\frac{\partial {T}}{\partial{x}} + {v}\frac{\partial {T}}{\partial{y}} \bigg) = \kappa_{hnf} \frac{ \partial^{2} T}{\partial y ^{2}} + q''' + \frac{\rho_{p} C_{s}}{\tau_{T}} (T_{p}-T)
\end{equation}
For dust particle phase:
\begin{equation}
\label{dim-contp}
\frac{\partial {u_{p}}}{\partial {x}}=-\frac{\partial {v_{p}}}{\partial {y}},
\end{equation}
\begin{equation}
\label{dim-v'}
u_{p} \frac{\partial {u_{p}}}{\partial {x}}+ v_{p} \frac{\partial {u_{p}}}{\partial {y}}=-\frac{K}{n}(u_{p}-u),
\end{equation}
\begin{equation}
\label{dim-t'}
u_{p} \frac{\partial {T_{p}}}{\partial{x}} + v_{p} \frac{\partial {T_{p}}}{\partial{y}} = -\frac{1}{\tau_{T}} (T_{p}-T).
\end{equation}
The citation for the equation describing nonlinear heat flux is given as \cite{1}.
\begin{equation}
q''' = \frac{k_{hnf}u_{w}(x)}{x \nu_{hnf}} [A (T_{w}-T_{\infty})f' + B (T-T_{\infty})],
\end{equation}
Regarding heat generation and absorption in fluid motion with $(A, B)>0$ and $(A, B)<0$, the modified energy equation (3) takes the following form:
\begin{equation*}
(\rho C_p)_{hnf} \bigg({u}\frac{\partial {T}}{\partial{x}} + {v}\frac{\partial {T}}{\partial{y}} \bigg) = \kappa_{hnf} \frac{ \partial^{2} T}{\partial y ^{2}} + \frac{k_{hnf}U_{w}}{x \nu_{hnf}}[A (T_{w}-T_{\infty})f' + B (T-T_{\infty})] 
\end{equation*} 
\begin{equation}
\label{dim_temp}
+ \frac{\rho_{p} C_{s}}{\tau_{T}} (T_{p}-T)
\end{equation}

subjected to boundary conditions given below (\cite{6} and \cite{7}):  
\begin{align*}
\label{dim_bc0}
&u=u_{w}(x)=U_{w}x^{m}, \,\,\,\ v=0, \,\,\ \kappa \frac{\partial {T}}{\partial{y}}= h_{f}(T_{w}-T) \qquad at \,\ y=0,
\end{align*}
\begin{equation}
\label{dim_bcinf}
u\longrightarrow u_{e}(x),\,\,\, u_{p} \longrightarrow 0,\,\,\, v_{p} \longrightarrow v,\,\,\, T\longrightarrow T_{\infty} ,\,\,\, T_{p} \longrightarrow T_{\infty}     \qquad  as \,\ y\longrightarrow\infty,
\end{equation}

where $(u,v, u_{p},v_{p})$ symbolizes the velocity constituents for the fluid and dust particle phases along the $(x,y)$ directions, respectively. $\kappa_{hnf}$, $\rho_{hnf}$, $\mu_{hnf}$, $\delta_{hnf}$, and $(\rho C_{p})_{hnf}$ stand for the thermal conductivity, density, dynamic viscosity, electrical conductivity, and specific heat capacitance of the hybrid nanofluid. The Stokes resistance parameter is represented by $K$, the dust particle density number by $N$, the thermal equilibrium time by $\tau_{t}$, the mass density of dust particles by $n$, the heat transfer parameter by $h_{f}$, the fluid temperature by $T$, the temperature of dust particles by $T_{p}$, the surface temperature by $T_{w}$, and the ambient temperature by $T_{\infty}$. Subsequently, a set of geometric transformations is applied to the system of equations (1-6) and (8-9) to facilitate the implementation of the numerical technique outlined in \cite{7}.
\begin{equation*}
\label{Transformation}
(u,u_{p})=\frac{\partial \psi}{\partial y}, \,\,\,\ (v,v_{p})=-\frac{\partial \psi}{\partial x},  \,\,\,\  \psi= \sqrt{\frac{2U_{e}x^{(m+1)}\nu_{f}}{(m+1)}}(f(\eta),f_{p}(\eta)),
\end{equation*}	
\begin{equation}
\label{nondd}
\eta=\sqrt{\frac{(m+1)x^{(m-1)}U_{e}}{2\nu_{f}}}y,  \,\,\,\ (\theta(\eta),\theta_{p}(\eta)) , =\frac{(T, T_{p})-T_{\infty}}{T_{w}-T_{\infty}}  
\end{equation}

The equations of continuity are fulfilled for both the fluid and particle phases, and the non-dimensional set of nonlinear ordinary differential equations for the fluid and dust phase are:
\begin{equation}
\label{nondim_f}
f''' + A_{1}\bigg[ff''-\Lambda f'^{2} + A_{2}(2-\Lambda)M\frac{\delta_{hnf}}{\delta_{f}}sin^{2}\omega f' + D_{\rho}\alpha_{d}H(f'-f'_{p})\bigg]=0 ,
\end{equation}
\begin{equation}
\label{nondim_theta}
f\theta'+ \frac{A_{3} K_{hnf}}{Pr K_{f}} \theta''+ (2-\Lambda) \bigg[ \frac{\lambda A_{3}K_{hnf}}{A_{1}PrK_{f}} (Af'+B\theta)+ Pr\Gamma \alpha_{d}(\theta_{p}-\theta) \bigg]=0,
\end{equation}
\begin{equation}
\label{nondim_f'}
f_{p}f_{p}''-\Lambda f_{p}'^{2} +\alpha_{d}(f'-f'_{p})=0 ,
\end{equation}
\begin{equation}
\label{nondim_theta'}
f_{p}\theta_{p}'+  (2-\Lambda) \alpha_{t}(\theta-\theta_{p})=0,
\end{equation}
and the resulting dimensionless boundary conditions are: 
\begin{align*}
\label{nondim_bc0}
f'(\eta)=\lambda,\,\,\  f(\eta)=0,\,\,\ \theta'(\eta)=\gamma(2-\Lambda)^{1/2}(1-\theta(\eta)) \qquad at \,\ \eta=0,
\end{align*}
\begin{equation}
\label{nondim_bcinf}
f'(\eta)=0,\,\,\  f'_{p}(\eta)=0, \,\,\  f_{p}(\eta)=f_{p}(\eta), \,\,\ \theta(\eta)=0, \,\,\ \theta_{p}(\eta)=0 \qquad at \,\ \eta \longrightarrow \infty
\end{equation}
Table \ref{tableNomenclature} contains the values of appearing parameters in the above equations. 

\begin{table}[h!]
	\caption{List of non-dimensional terms appearing in Equations (\ref{nondim_f}-\ref{nondim_bcinf})}
 \label{tableNomenclature}
	\begin{center}
		\begin{tabular}{|c c|} 
			\hline
$ D_{\rho}= \frac{mN}{\rho_{p}} $  & Mass concentration of dust particles  \\
$\alpha_{d}= \frac{2x}{\tau_{v}(m+1)U_{e}x^{m}}$  &  particle interaction parameter\\
$ H=\frac{\rho_{p}} {\rho_{hnf}} $  & Relative density  \\
$ \Gamma= \frac{C_{s}}{(C_{p})_{hnf}} $ &  Ratio of specific heat \\
$ \alpha_{t}= \frac{x^{(1-m)}}{\tau_{t}U_{e}} $ &  Fluid particle interaction parameter   \\
$ \lambda= \frac{U_{w}}{U_{e}} $  &  Velocity ratio parameter \\
$ Pr= \frac{(\rho C_{p})_{f}\nu_{f}}{\kappa_{f}} $ &  Prandtl number \\
$ M= \frac{\beta_{0}^{2}x^{(1-m)}\delta_{f}}{\rho_{f}U_{e}} $   & Magnetic parameter \\
$  \gamma= \frac{h_{f}(\nu_{f}x^{1-m}/U_{e})^{1/2}}{K} $  & Dimensionless Biot number  \\
$\Lambda= \frac{2m}{m+1}$ &  Wedge angle parameter\\
$ A_{1}=(1-\phi)^{2.5}[(1-\phi)+\phi\frac{\rho_{s}}{\rho_{f}}] $  & \\ 
$ A_{2}=\frac{\rho_{f}}{\rho_{f}(1-\phi)+\rho_{s}\phi} $  &  \\
$ A_{3}=\frac{(\rho C_{p})_{f}}{(\rho C_{p})_{hnf}} $ &  \\ 
            \hline
		\end{tabular}
	\end{center}
\end{table}

\section{Material's Thermophysical Properties}

The data provided in Table \ref{table1} delineates the traits of the base fluid $(EG)$ and the nanoparticles $(Cu + SiO_2)$. A notable study conducted by \cite{9} suggests that nanofluids exhibit non-Newtonian fluid behavior when the volume fraction of nanoparticles exceeds 5-6\%. In alignment with this finding, the current study adopts a similar approach, exploring volume fractions of $Cu$ and $SiO_2$ nanoparticles ranging from 0 to 0.1\%, consistent with the work of \cite{9}. The thermal characteristics of the nanofluid and relevant formulas are outlined in Table \ref{table2}, where $\phi$ represents the volume fraction of nanoparticles, $\rho_{f}$ is the dynamic viscosity of the base fluid, $\rho_{s}$ is the dynamic viscosity of solid particles, $\mu_{f}$ is the dynamic viscosity of the fluid, $(\rho C_{p}){f}$ is the heat capacitance of the fluid, $(\rho C{p}){s}$ is the thermal capacitance of particles, $\kappa{f}$ is the thermal conductivity of the fluid, $\kappa_{s}$ is the thermal conductivity of particles, $\delta_{f}$ is the electrical conductivity of the fluid, and $\delta_{s}$ is the electrical conductivity of particles.

The enhancement of heat transfer rates can be achieved by employing a hybrid nanofluid, as it exhibits a higher heat exponent compared to traditional nanofluids \cite{p1, p2, p3}. The pertinent variables for the hybrid nanofluid are presented in Table \ref{table3}, where $\mu_{hnf}, \rho_{hnf}, (\rho C_{p}){hnf}, \kappa{hnf}, \delta_{hnf}$ denote the dynamical viscosity, consistency, specific-heat capacitance, thermal and electrical conductance of the hybrid nanofluid. Similarly, $\mu_{f}, \rho_{f}, (C_{p}){f}, \kappa{f}, \delta_{f}$ represent the dynamical viscosity, consistency, specific heat capacitance, thermal, and electrical conductance of the base liquid. Additionally, $\rho_{p1}, \rho_{p2}, (C_{p}){p1}, (C{p}){p2}, \kappa{p1}, \kappa_{p2}, \delta_{p1}, \delta_{p2}$ signify the consistency, specific heat capacitance, thermal and electrical conductance of the solid particles. The parameter $\phi$ stands for the volume accumulation coefficient of nanosolid particles for conventional nanofluids, while $\phi_{hnf}=\phi_{Cu}+\phi_{SiO_{2}}$ represents the solid-particle concentration factor for the mixture nanofluid \cite{p4, p5, p6, p7}.
\begin{table}[h!]
	\caption{Thermophysical properties of ordinary fluid and nanoparticles \cite{8} } 
\label{table1}
 \begin{center}
		\begin{tabular}{|c c c c|} 
			\hline
			Thermophysical Properties & Copper $(Cu)$ & Silicon Dioxide $(SiO_{2})$ & Ethylene Glycol $(CH_{2}OH)_{2}$ \\ [0.5ex] 
			\hline
			$\rho /kg m^{-3}$ & 8933 & 2650 & 1110 \\
   
                 $C_{p}/J Kg^{-1}$ & 385 & 730 & 22000 \\ 
			
			$ \kappa /W m K^{-1}$ & 401 & 1.5 & 0.253 \\
			
			$\sigma/ \Omega m^{-1}$  & $5.96 \times 10^{7}$ & $1.0 \times 10^{-18}$ & $5.5 \times 10^{-6}$ \\
			
			\hline
		\end{tabular}
	\end{center}
\end{table}
\begin{table}[h!]
	\caption{Expressions for thermophysical properties of nanofluids} 
 \label{table2}
	\begin{center}
		\begin{tabular}{|c c|} 
			\hline
			Properties & Nanofluid \\ [0.5ex] 
			\hline
			Dynamic viscosity $(\mu)$ & $\mu_{nf} = \mu_{f}(1-\phi)^{-2.5}$ \\
   
                Thermal conductivity $(\kappa)$ & $\frac{\kappa_{nf}}{\kappa_{f}} = \bigg[\frac{(\kappa_{s}+2\kappa_{f})-2\phi(\kappa_{f}-\kappa_{s})} {(\kappa_{s}+2\kappa_{f})+\phi(\kappa_{f}-\kappa_{s)}}\bigg] $ \\ 
			
			Electrical conductivity $(\delta)$ & $\frac{\delta{nf}}{\delta{f}} = \bigg[1+\frac{3\big(\frac{\delta_{s}}{\delta_{f}}-1 \big)\phi}{\big(\frac{\delta_{s}}{\delta_{f}}+2 \big)-\big(\frac{\delta_{s}}{\delta_{f}}-1 \big)\phi}  \bigg]$  \\
			
			Density $(\rho)$ & $\rho_{nf} = \rho_{f}(1-\phi) - \rho_{s}\phi $ \\

                Heat capacity $(\rho C_{p})$ & $(\rho C_{p})_{nf} = (\rho C_{p})_{f}(1-\phi)-(\rho C_{p})_{s}\phi$ \\
			\hline
		\end{tabular}
	\end{center}
\end{table}

\begin{table}[h!]
\caption{Expressions for thermophysical properties of  hybrid nanofluids}
\label{table3}	
 \begin{center}
\begin{tabular}{|c c|} 
\hline
Properties & Hybrid nanofluid \\ [0.5ex] 
\hline
Dynamic viscosity $(\mu)$ & $\mu_{hnf} = (1-\phi_{Cu})^{-2.5}(1-\phi_{SiO_{2}})^{-2.5}\mu_{f}$ \\
 Thermal conductivity $(\kappa)$ & $\frac{\kappa_{hnf}}{\kappa_{gf}} = \bigg[\frac{(\kappa_{p2}+2\kappa_{gf})-2\phi_{SiO_{2}}(\kappa_{gf}-\kappa_{p2})} {(\kappa_{p2}+2\kappa_{gf})+\phi_{SiO_{2}}(\kappa_{gf}-\kappa_{p2)}}\bigg] $ , $\frac{\kappa_{gf}}{\kappa_{f}} = \bigg[\frac{(\kappa_{p1}+2\kappa_{f})-2\phi_{Cu}(\kappa_{f}-\kappa_{p1})} {(\kappa_{p1}+2\kappa_{gf})+\phi_{Cu}(\kappa_{f}-\kappa_{p1)}}\bigg] $ \\ 
Electrical conductivity $(\delta)$ & $\frac{\delta{nf}}{\delta{f}} = \Bigg[1+ \frac{3\big(\frac{\phi_{Cu}\delta_{p1}+\phi_{SiO_{2}}\delta_{p2}}{\delta_{f}}-(\phi_{Cu}+\phi_{SiO_{2}}) \big)}{\big(\frac{\phi_{Cu}\delta_{p1}+\phi_{SiO_{2}}\delta_{p2}}{(\phi_{Cu}+\phi_{SiO_{2}})\delta_{f}} +2\big)-\big(\frac{\phi_{Cu}\delta_{p1}+\phi_{SiO_{2}}\delta_{p2}}{\delta_{f}} -(\phi_{Cu}+\phi_{SiO_{2}})\big)} \Bigg]$  \\
Density $(\rho)$ & $\rho_{hnf} = [(1-\phi_{SiO_{2}})((1-\phi_{Cu})\rho_{f}+\phi_{Cu}\rho_{p1})] + \phi_{SiO_{2}}\rho_{p2} $ \\
 Heat capacity $(\rho C_{p})$ & $(\rho C_{p})_{hnf} = [(1-\phi_{SiO_{2}})((1-\phi_{Cu})(\rho C_{p})_{f}+\phi_{Cu}(\rho C_{p})_{p1})]+\phi_{SiO_{2}}(\rho C_{p})_{p2}$ \\
\hline
\end{tabular}
\end{center}
\end{table}

\section{Physical quantities of interest}

Two significant physical parameters in fluid dynamics are the skin friction or drag coefficient and the local Nusselt number. The drag coefficient is a measure of the frictional drag force experienced by a fluid flowing over a surface, while the local Nusselt number is a dimensionless parameter that indicates the rate of heat transfer at a given point on the surface. These quantities are crucial for understanding the behavior of fluids in various applications, including aerospace, automotive engineering, and thermodynamics and are defined by specific expressions: 

Skin friction coefficient:
\begin{align*}
    C_{f} = \frac{\tau_{w}}{\rho_{f} u_{e}^2} ,\,\,\,\,\,\  \tau_{w} = \mu_{hnf} \bigg( \frac{\partial u}{\partial y} \bigg)_{y = 0},
\end{align*}
 and local Nusselt number:
\begin{align}
    Nu_{x} = \frac{x q_{w}}{k_{f} (T_{w} - T_{\infty})} ,\,\,\,\,\,\  q_{w} = -k_{hnf} \bigg( \frac{\partial T}{\partial y} \bigg)_{y = 0},
\end{align}
The non-dimensionalized form for the aforementioned expressions can be attained through the utilization of equation (\ref{nondd}).
\begin{align*}
   \sqrt{\frac{2 Re}{m+1}} C_{f} = \frac{f''(0)}{(1-\phi_{Cu})^{2.5}(1-\phi_{SiO_{2}})^{2.5}} ,
\end{align*}
 and 
 \begin{align}
     \sqrt{\frac{2}{(m+1)Re} } Nu_{x} =- \frac{k_{hnf}}{k_{f}} \theta'(0) ,
\end{align}

\section{Numerical Methodology}

In this section, we elucidate the numerical methodology employed for solving the system of non-dimensional ordinary differential equations ($\ref{nondim_f}$)-($\ref{nondim_bcinf}$), derived from the partial differential equations ($\ref{dim_continuity}$)-($\ref{dim_bcinf}$), which characterize the temporal evolution of the non-dimensional variables. The numerical solution was obtained using the bvp4c MATLAB built-in function, a robust and widely utilized numerical integration technique \cite{n1, n8}. This method employs a collocation approach with a fourth-order Lobatto IIIa formula, offering a precise and efficient means of solving boundary value problems (BVPs). Known for its simplicity and high accuracy, bvp4c is favored in various scientific and engineering applications. The function accommodates the variables in a specific form, ensuring effective handling and solution of the given set of dimensionless ordinary differential equations.

\begin{align*}
f = k_{1}, \,\,\ f' = k_{2}, \,\,\ f'' = k_{3}, \,\,\ f_{p} = k_{6}, \,\,\ f'_{p} = k_{7}, \,\,\ \theta = k_{4}, \,\,\ \theta' = k_{5}, \,\,\ \theta_{p} = k_{8},
\end{align*}
Based on the above equations, the following equations can be obtained using MATLAB bvp4c:
\begin{align*}
k_{3}' = -A1\bigg[k_{1}k_{3} - \beta k_{2}^{2} + A2(2-\beta)M\frac{\delta_{hnf}}{\delta_{f}}sin^{2}\omega(1-k_{2})+D_{\rho}\alpha_{d}H(k_{2}-k_{7})\bigg],
\end{align*}
\begin{align*}
k_{5}' = -\frac{pr\kappa_{f}}{A_{3}\kappa_{hnf}} \bigg[k_{1}k_{5} + (2-\beta) \frac{\lambda A3\kappa_{hnf}}{prA1\kappa_{f}} (Ak_{2}+Bk_{4}) + pr\Gamma\alpha_{t}(k_{8}-k_{4}) \bigg],
\end{align*}
\begin{align*}
k_{6}k_{7}' = \bigg[\beta k_{7}^{2} - \alpha_{d} (k_{2}-k_{7})\bigg],
\end{align*}
\begin{align*}
k_{6}k_{8}' = (2-\beta)\alpha_{t} (k_{8}-k_{2}),
\end{align*}
and boundary conditions become:
\begin{align*}
k_1(0) = 0, \,\,\,\ k_2(0)-\lambda = 0, \,\,\,\ k_5(0)+\gamma (2-\beta)^{1/2}(k_4(0)-1) = 0,  \,\,\,\ k_7(0) = 0,  \,\,\,\ k_6(0)- k_1(0)= 0,
\end{align*}
\begin{align*}
k_1(\infty)-1= 0, \,\,\,\ k_4(\infty)= 0, \,\,\,\ k_7(\infty)= 0, \,\,\,\ k_8(\infty)= 0
\end{align*}
The bvp4c solver, through MATLAB script, iteratively employs a fourth-order collocation method, which means it approximates the solution using cubic polynomials that satisfy the differential equation at four points (the collocation points) in each subinterval of a mesh. It automatically adjusts the mesh to satisfy the error tolerances up to $10^{-6}$.

\section{Findings and Discussions}
In this section, a comprehensive graphical analysis of the physical aspects encompassing momentum, energy, and all relevant parameters is presented through Figures (\ref{f-1})-(\ref{f-19}). The numerical computation of the dimensionless boundary value problem (BVP) ($\ref{nondim_f}$)-($\ref{nondim_bcinf}$) and the physical properties characterizing the flow of dusty hybrid nanofluid ($\ref{nondd}$) is conducted through the bvp4c MATLAB built-in method. The obtained results are visually depicted in the figures, considering various influential parameters such as $\phi_{1}=\phi_{2}=0.00, 0.05$, $A=B=-0.5, 0.5$, $\lambda=0.1, 0.2, 0.5$, $M=1.0, 2.0, 3.0$, $D_{\rho}=\alpha_{d}=0.2$, $m=1.0$, $H=0.1$, and $\omega=\pi/2$. The straight red lines denote the first solution, while the dotted green lines represent the second solution. The graphical representations elucidate the impacts of these determinants on the traditional nanofluid, hybrid nanofluid, and dusty hybrid nanofluid flow over both horizontal and vertical stretching plates of a wedge body.

\subsection{Inclinated magnetic field parameter and velocity variation}
\begin{figure}[h]
\centering
\includegraphics[width=13cm, height=5.2cm]{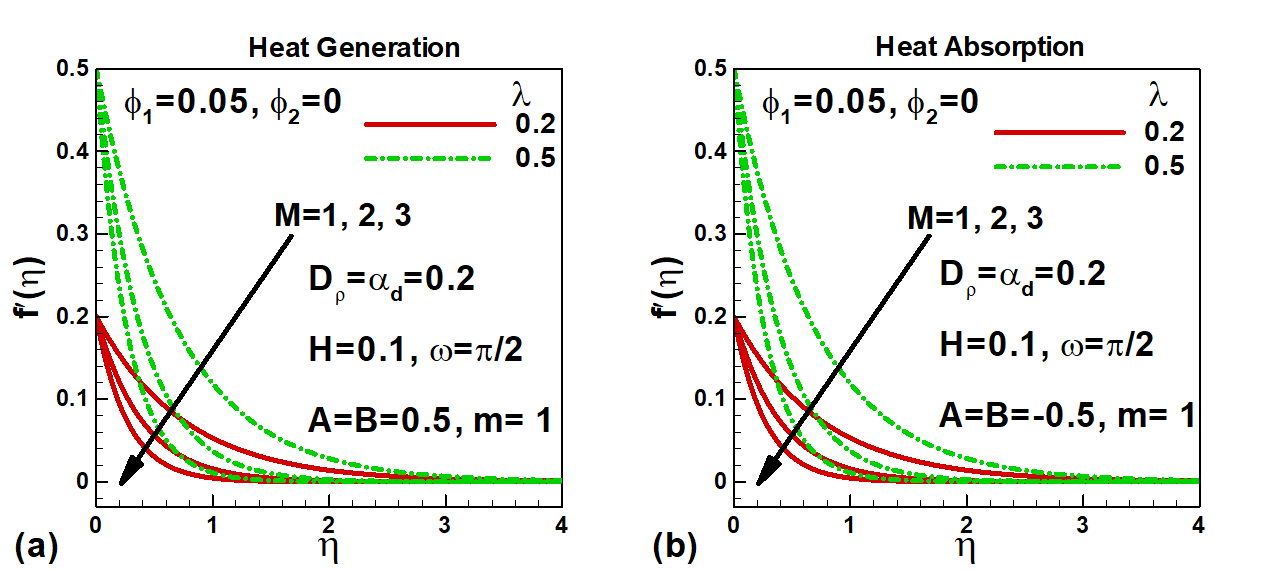}
\caption{Effects of inclined magnetic field on the dimensionless velocity of nanofluid near the plane stagnation point on a vertical stretching flat plate with heat (a) generation and (b) absorption}
\label{f-1}
\end{figure}

\begin{figure}[h]
\centering
\includegraphics[width=13cm, height=5.2cm]{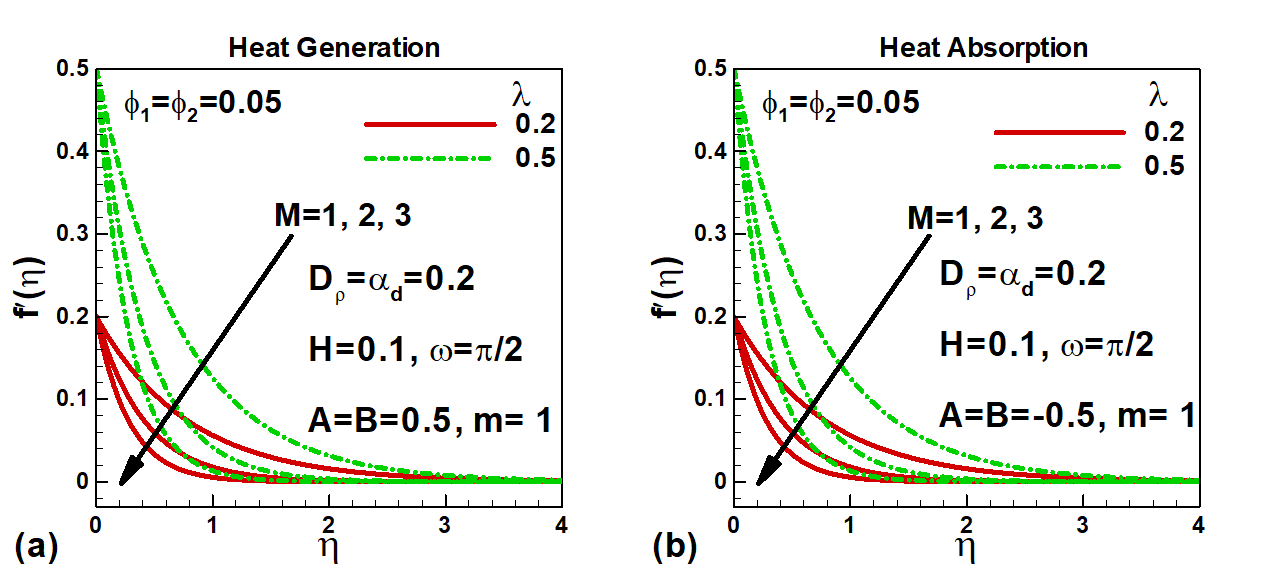}
\caption{Effects of inclined magnetic field on dimensionless velocity of hybrid nanofluid near plane stagnation point on vertical stretching flat plate with heat (a) generation and (b) absorption}
\label{f-2}
\end{figure}

\begin{figure}[h]
\centering
\includegraphics[width=13cm, height=5.2cm]{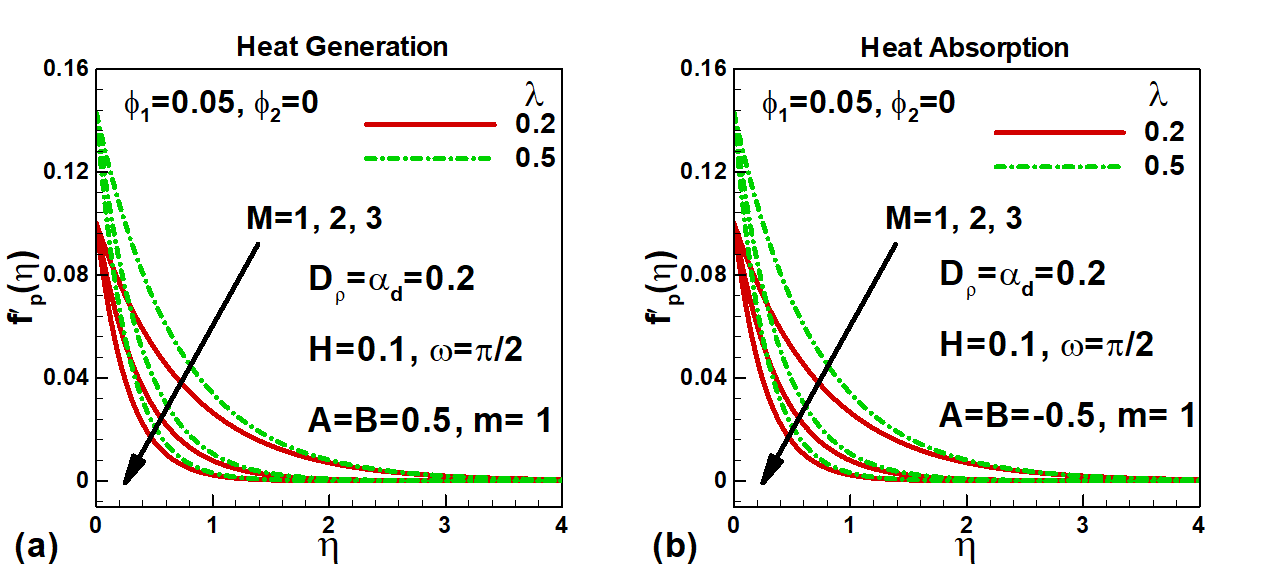}
\caption{Effects of inclined magnetic field on dimensionless velocity of dusty nanofluid near plane stagnation point on vertical stretching flat plate with heat (a) generation and (b) absorption}
\label{f-3}
\end{figure}

\begin{figure}[h]
\centering
\includegraphics[width=13cm, height=5.2cm]{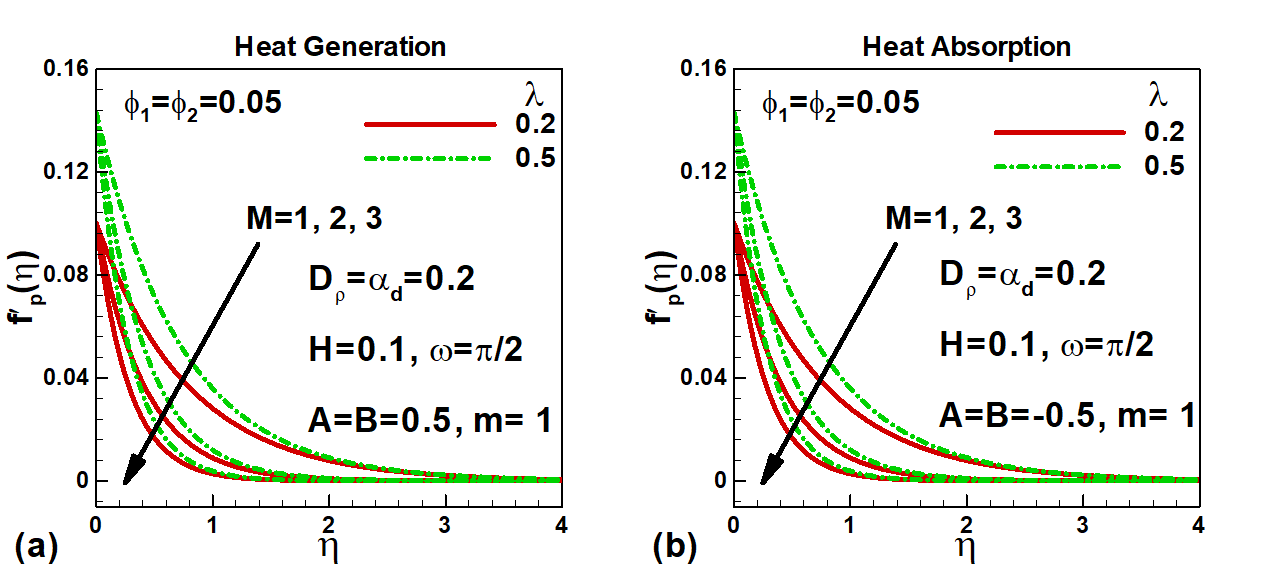}
\caption{Effects of inclined magnetic field on dimensionless velocity of dusty hybrid nanofluid near plane stagnation point on vertical stretching flat plate with heat (a) generation and (b) absorption}
\label{f-4}
\end{figure}

The plots (\ref{f-1}-\ref{f-4}) present a detailed analysis of the influence of an inclined magnetic force on the velocity profile concerning varying volume fractions of added particles. Figures (\ref{f-1}) and (\ref{f-2}) consistently illustrate that an escalation in the magnetic field $(M)$ is affiliated with diminishing trend in the dimensionless velocity of both the nanofluid and hybrid nanofluid near the stagnation point. This reduction in velocity with increasing $M$ holds true for scenarios involving both heat generation and absorption. The magnetic field, operating through the nonlinear heat flux mechanism, exerts a suppressive effect on fluid motion, leading to decreased velocity profiles for both the nanofluid and hybrid nanofluid. The application of a magnetic field to a conductive fluid introduces the Lorentz force, which serves as a resistance force perpendicular to flow direction and magnetic force. Consequently, this Lorentz force opposes the motion of the fluid, contributing to the observed reduction in velocity.
The variations in the magnetic field against the  dusty nanofluid and hybrid dusty nanofluid velocity near the plane stagnation point on a vertically stretching flat plate are depicted in figure (\ref{f-3}) and figure (\ref{f-4}). A noticiable trend indicates that the momentum profiles of both dusty nanofluid and hybrid dusty nanofluid exhibit a decline with increasing magnetic field strength. This phenomenon can be attributed to the nonlinear heat flux associated with the magnetic field, which tends to dampen the thermal and fluid dynamic processes within the system. The damping effect plays a crucial role in reducing fluid velocity by dissipating energy within the system. As the magnetic field strength intensifies, this damping effect becomes more pronounced, ultimately resulting in a significant decrease in fluid velocity. 

\subsection{Inclinated magnetic field and Temperature variation}

\begin{figure}[h]
\centering
\includegraphics[width=13cm, height=5.2cm]{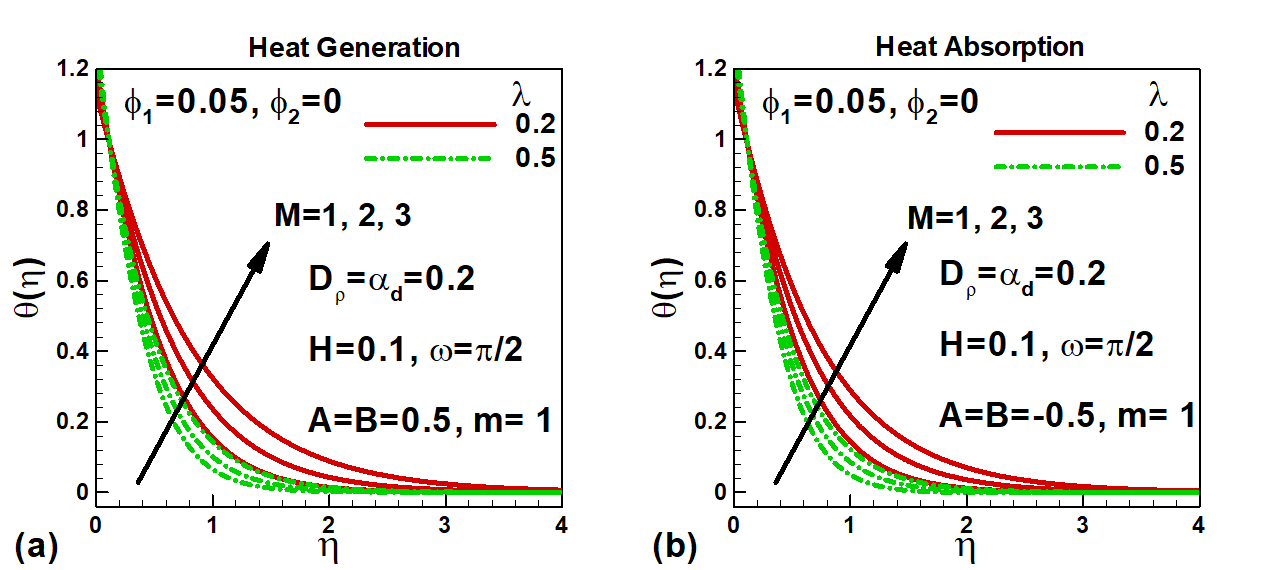}
\caption{Effects of inclined magnetic field on dimensionless temperature of nanofluid near plane stagnation point on vertical stretching flat plate with heat (a) generation and (b) absorption}
\label{f-5}
\end{figure}

\begin{figure}[h]
\centering
\includegraphics[width=13cm, height=5.2cm]{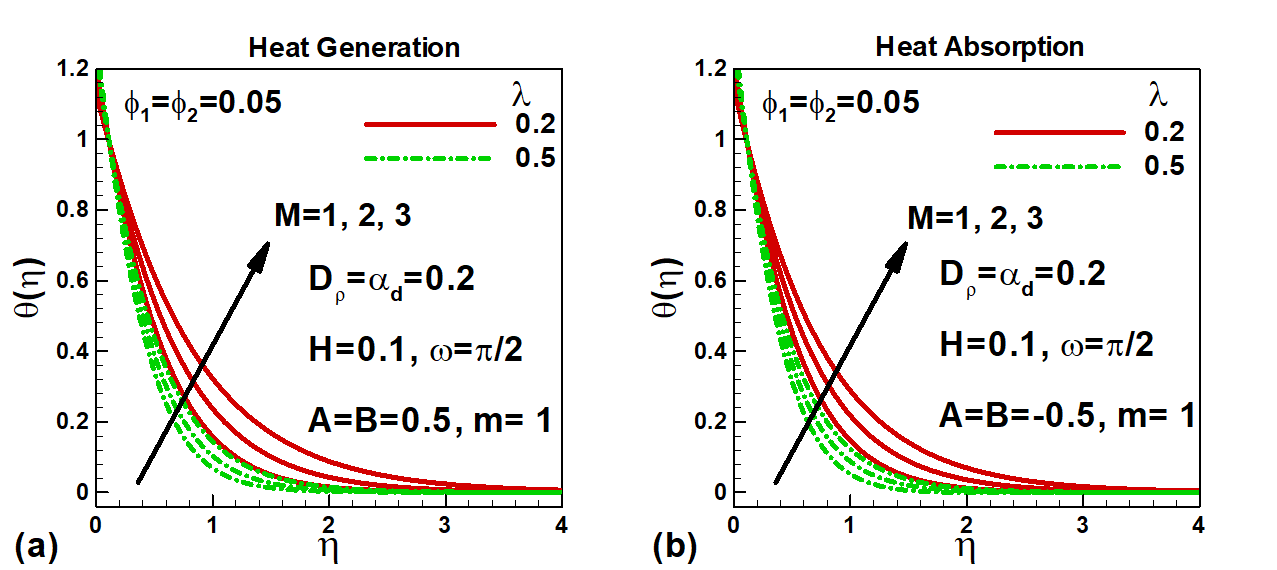}
\caption{Effects of inclined magnetic field on dimensionless temperature of hybrid nanofluid near plane stagnation point on vertical stretching flat plate with heat (a) generation and (b) absorption}
\label{f-6}
\end{figure}

\begin{figure}[h]
\centering
\includegraphics[width=13cm, height=5.2cm]{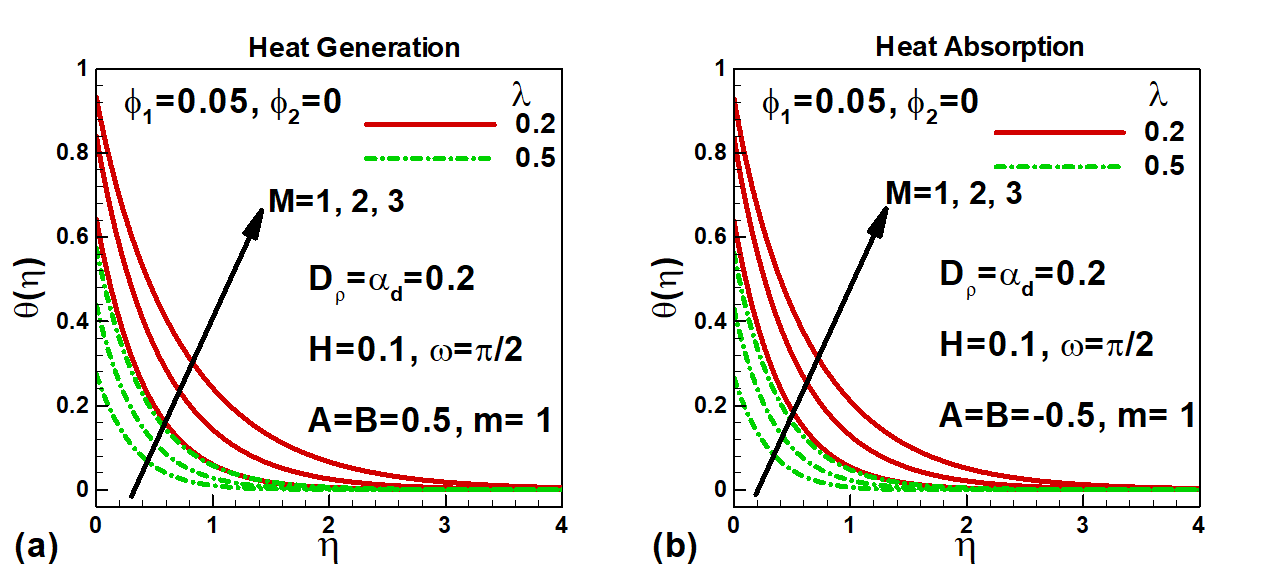}
\caption{Effects of inclined magnetic field on dimensionless temperature of dusty nanofluid near plane stagnation point on vertical stretching flat plate with heat (a) generation and (b) absorption}
\label{f-7}
\end{figure}

\begin{figure}[h]
\centering
\includegraphics[width=13cm, height=5.2cm]{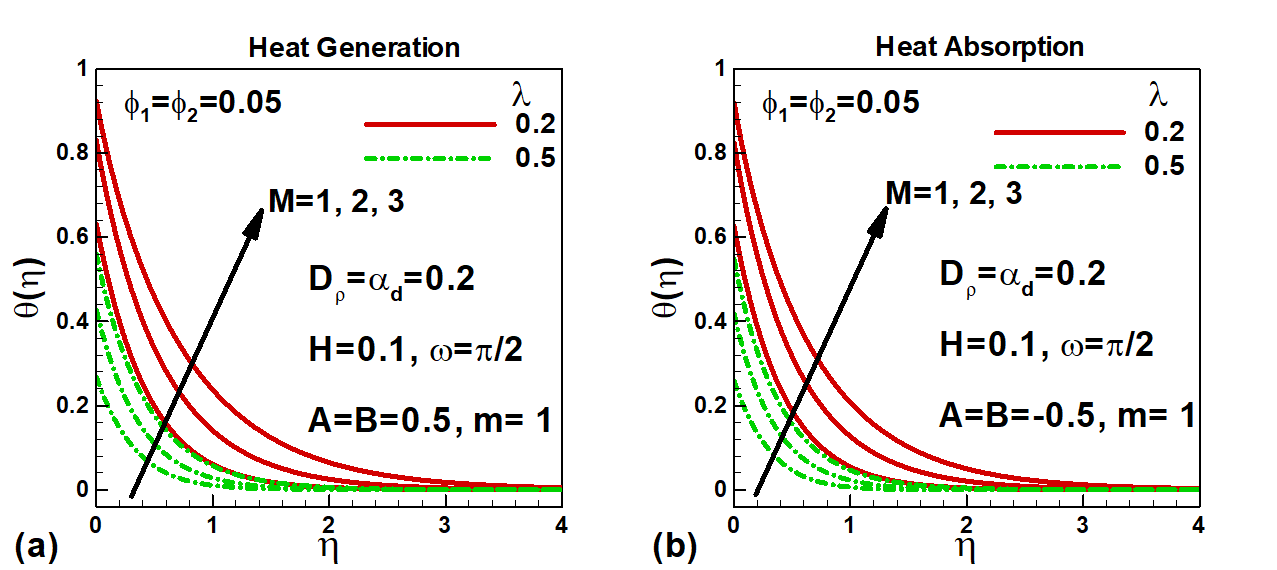}
\caption{Effects of inclined magnetic field on dimensionless temperature of dusty hybrid nanofluid near plane stagnation point on vertical stretching flat plate with heat (a) generation and (b) absorption}
\label{f-8}
\end{figure}

\begin{figure}[h]
\centering
\includegraphics[width=13cm, height=5.2cm]{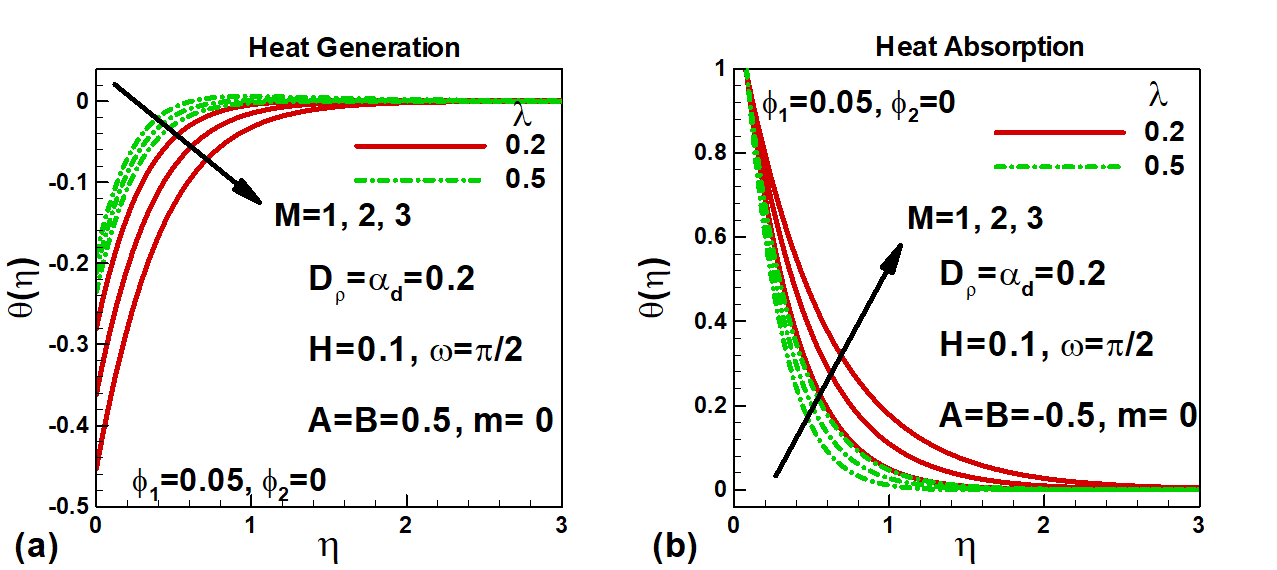}
\caption{Effects of inclined magnetic field on dimensionless temperature of nanofluid on horizontal stretching flat plate with heat (a) generation and (b) absorption}
\label{f-10}
\end{figure}

\begin{figure}[h]
\centering
\includegraphics[width=13cm, height=5.2cm]{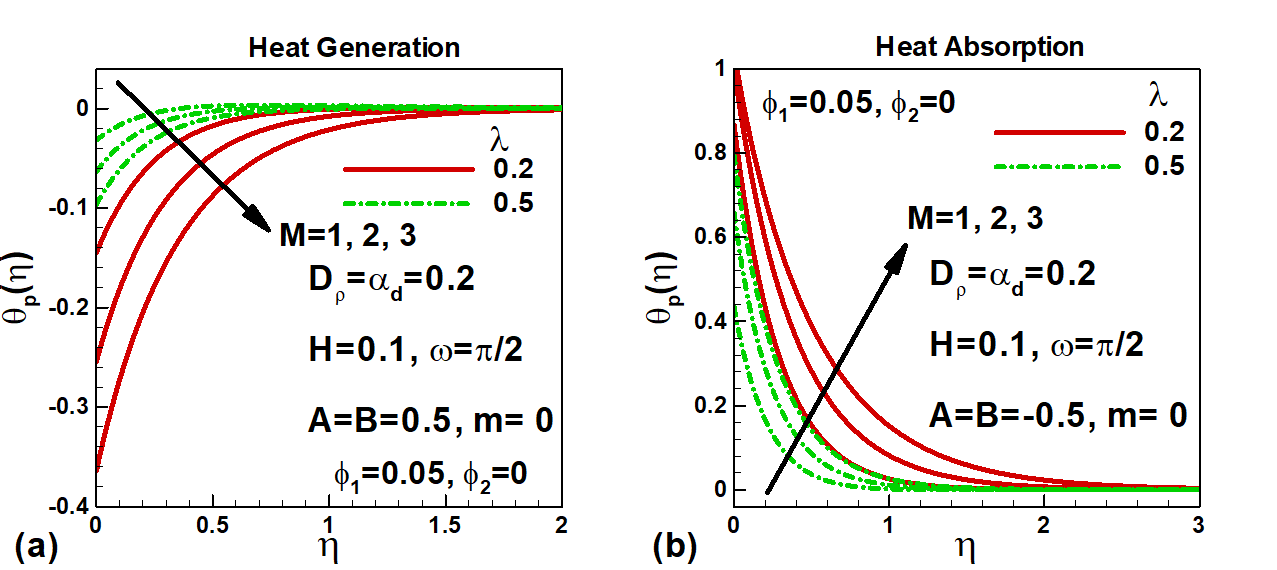}
\caption{Effects of inclined magnetic field on dimensionless temperature of dusty nanofluid on horizontal stretching flat plate with heat (a) generation and (b) absorption}
\label{f-11}
\end{figure}

The plots (\ref{f-5}-\ref{f-10}) provide insights into the influence of an inclined magnetic field on the momentum profile for varying volume fractions of added particles. The effect of this magnetic field on the dimensionless heat flow $\theta(\eta)$ for both nanofluids and hybrid nanofluids near both vertical and horizontal stretching flat plates is illustrated in figures (\ref{f-5}) and (\ref{f-10}). In figure (\ref{f-5}), where $\phi_{1} = 0.05$ and $\phi_{2} = 0$ is set for nanofluid, the temperature profile is observed to increase with the magnetic force $M$ for both heat generation and absorption across the studied range of $M=1, 2, 3$. The development in the convective boundary layer thickness is represented by steeper gradients of $\theta(\eta)$ for higher values of $M$. Figure (\ref{f-6}) explores the same phenomenon with $\phi_{1} = \phi_{2} = 0.05$ for hybrid nanofluid, indicating that the addition of $SiO_2$ nanoparticles influences the temperature profiles. In both cases of heat generation and absorption, the temperature profiles ascend with increasing $M$, and this trend remains consistent across different velocity ratio parameters $(\lambda = 0.2, 0.5)$. 
Figures (\ref{f-7}) and (\ref{f-8}) delve into the thermal effects of an inclined magnetic field on dusty nanofluids and hybrid nanofluids around a wedge body. In both figures, the dimensionless temperature, displays an growth with the magnetic field parameter $M$. Figure (\ref{f-7}) elucidates the behavior of a dusty nanofluid having a volume fraction $\phi_{1}$ of Ethylene glycol and $Cu$, while Figure (\ref{f-8}) focuses on a hybrid nanofluid with equal volume fractions $\phi_{1}$ and $\phi_{2}$ of Ethylene glycol, $Cu$, and $SiO_2$. The observed rise in temperature profiles with an increase in $M$ suggests that the magnetic force amplifies the energy profile, likely attributed to the impact of the induced Lorentz force on flow motion. For both heat generation and absorption scenarios, higher values of $M$ are associated with more pronounced convective boundary layers, indicating that the magnetic field's impact is more significant than the thermal buoyancy effects under the studied conditions. 
In Figures (\ref{f-10}) and (\ref{f-11}), the dimensionless temperature profiles for dusty and hybrid nanofluids are scrutinized under the influence of an inclined magnetic field characterized by the parameter $M$. In both figures, the dusty nanofluid (\ref{f-10}a) and the hybrid nanofluid (\ref{f-11}a) exhibit a decrease in the temperature profile with an increase in $M$ during heat generation. Conversely, during heat absorption (\ref{f-10}b) and (\ref{f-11}b), there is an improvement in the thermal profile with an increase in $M$. This observation indicates that the magnetic field effectively suppresses the development of the thermal boundary layer during heat generation while enhancing thermal energy dispersion during heat absorption. The results demonstrate the significant control exerted by $M$ over the thermal behavior of the fluids, emphasizing its dual role in either dampening or amplifying the thermal effects depending on the specific heat generation or absorption conditions. 
\subsection{Physical quantities against $M$ and $\lambda$}
\begin{figure}[h]
\centering
\includegraphics[width=13cm, height=5.2cm]{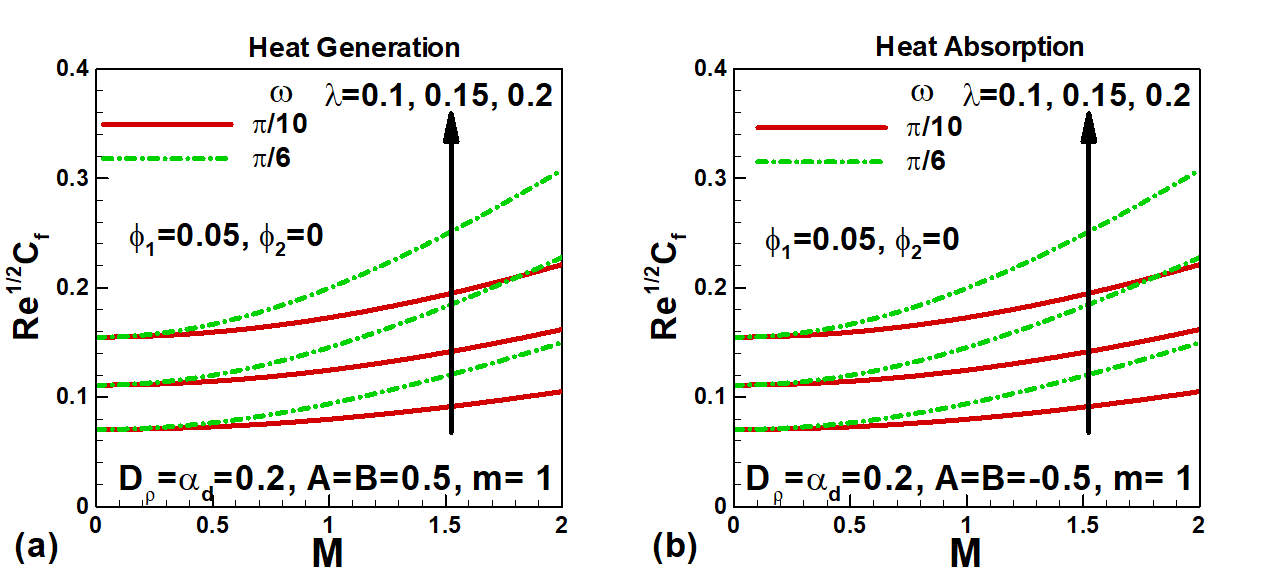}
\caption{Variation in skin friction with an inclined magnetic field for nanofluid near plane stagnation point on vertical stretching flat plate with heat (a) generation and (b) absorption}
\label{f-12}
\end{figure}

\begin{figure}[h]
\centering
\includegraphics[width=13cm, height=5.2cm]{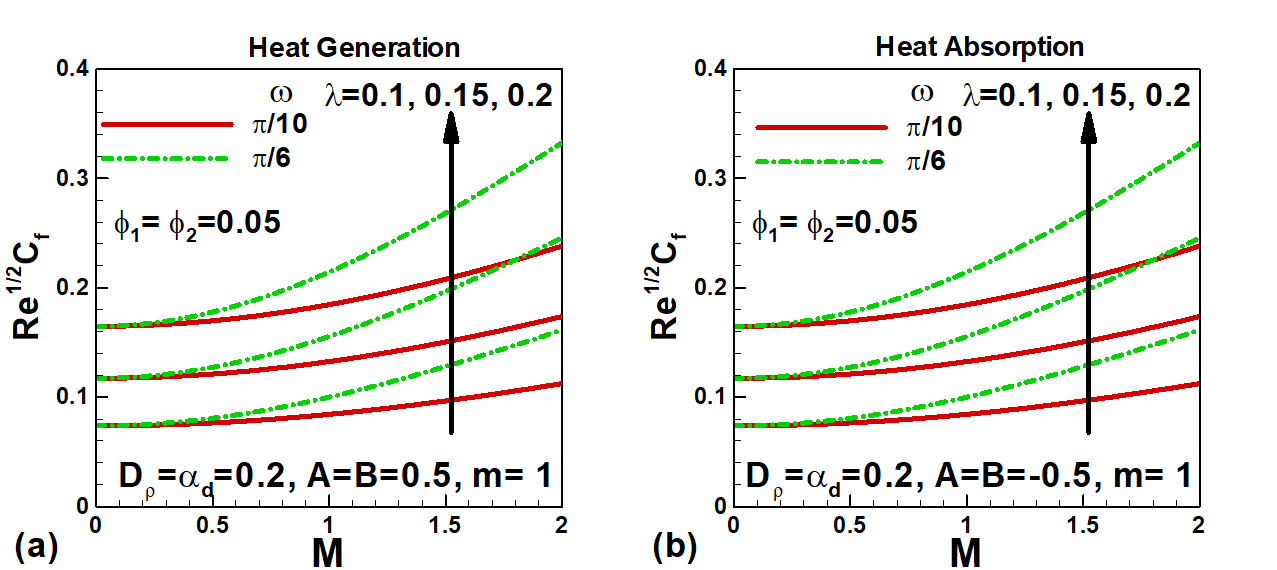}
\caption{Variation in skin friction with an inclined magnetic field for hybrid nanofluid near plane stagnation point on vertical stretching flat plate with heat (a) generation and (b) absorption}
\label{f-13}
\end{figure}

The plots (\ref{f-5}-\ref{f-10}) provide a comprehensive analysis of the impact of an inclined magnetic field on the velocity profile for varying volume fractions of added particles. Figures (\ref{f-12}) and (\ref{f-13}) focus on the skin friction coefficient $Re^2 C_{f}$ on a vertical stretching flat plate in a wedge flow concerning the velocity ratio parameter $\lambda$ for both nanofluids and hybrid nanofluids, with and without heat transfer. The results reveal that an increase in $\lambda$ corresponds to a rise in the skin friction for both heat generation and absorption, indicating a more substantial interaction between the flow and the wedge's surface as the momentum ratio escalates. This effect is consistent for both nanofluids (\ref{f-12}) and hybrid nanofluids (\ref{f-13}), demonstrating the sensitivity of the skin friction to changes in $\lambda$. 

\begin{figure}[h]
\centering
\includegraphics[width=13cm, height=5.2cm]{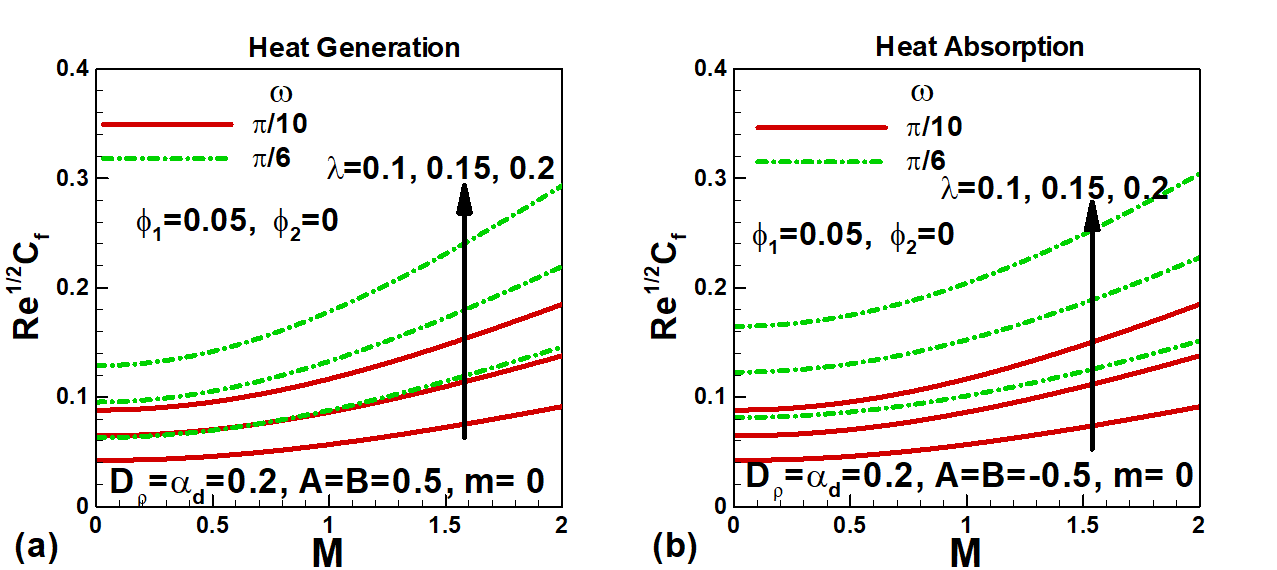}
\caption{Variation in skin friction with an inclined magnetic field for nanofluid on horizontal stretching flat plate with heat (a) generation and (b) absorption}
\label{f-14}
\end{figure}

\begin{figure}[h]
\centering
\includegraphics[width=13cm, height=5.2cm]{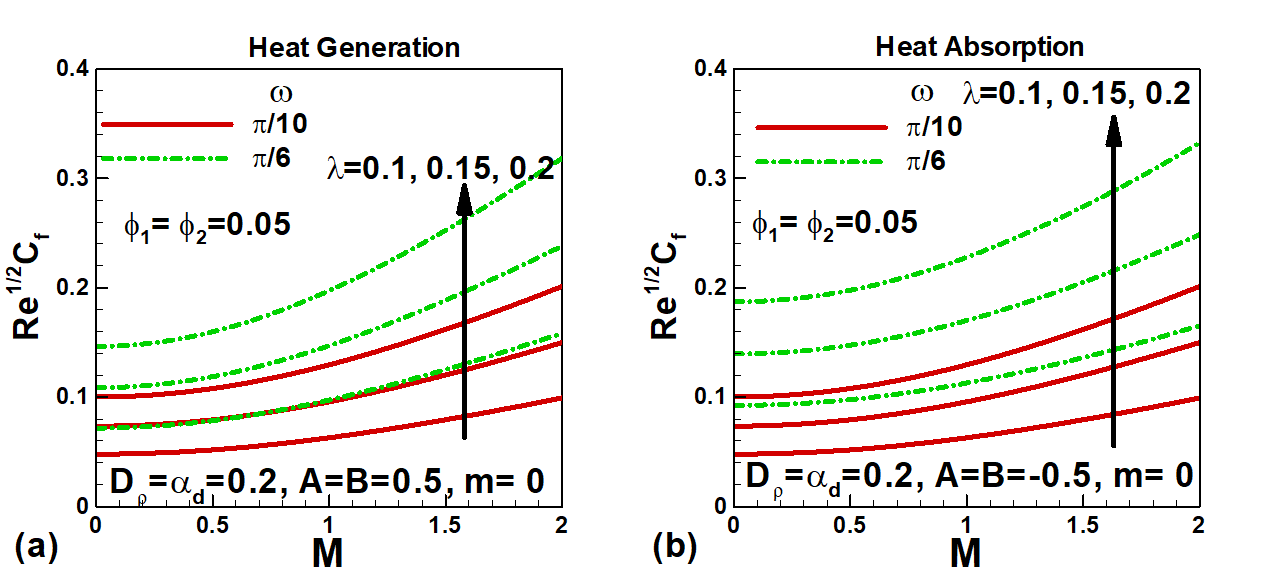}
\caption{Variation in skin friction with an inclined magnetic field on horizontal stretching flat plate with heat (a) generation and (b) absorption}
\label{f-15}
\end{figure}

\begin{figure}[h]
\centering
\includegraphics[width=13cm, height=5.2cm]{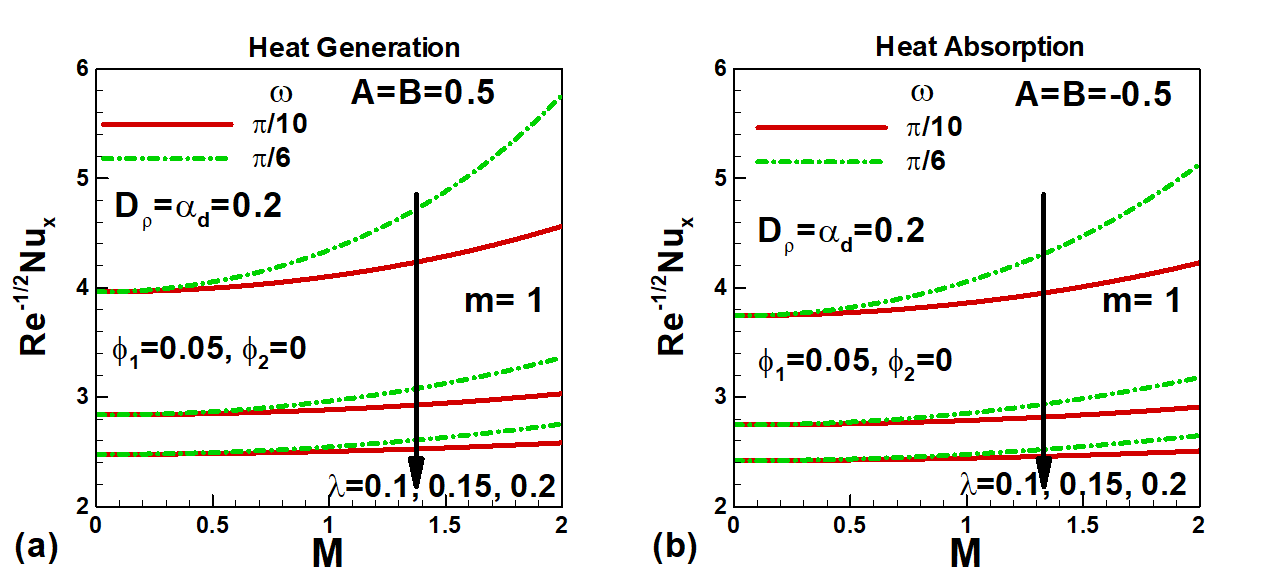}
\caption{Variation in Nusselt number with an inclined magnetic field for nanofluid near plane stagnation point on vertical stretching flat plate with heat (a) generation and (b) absorption}
\label{f-16}
\end{figure}

\begin{figure}[h]
\centering
\includegraphics[width=13cm, height=5.2cm]{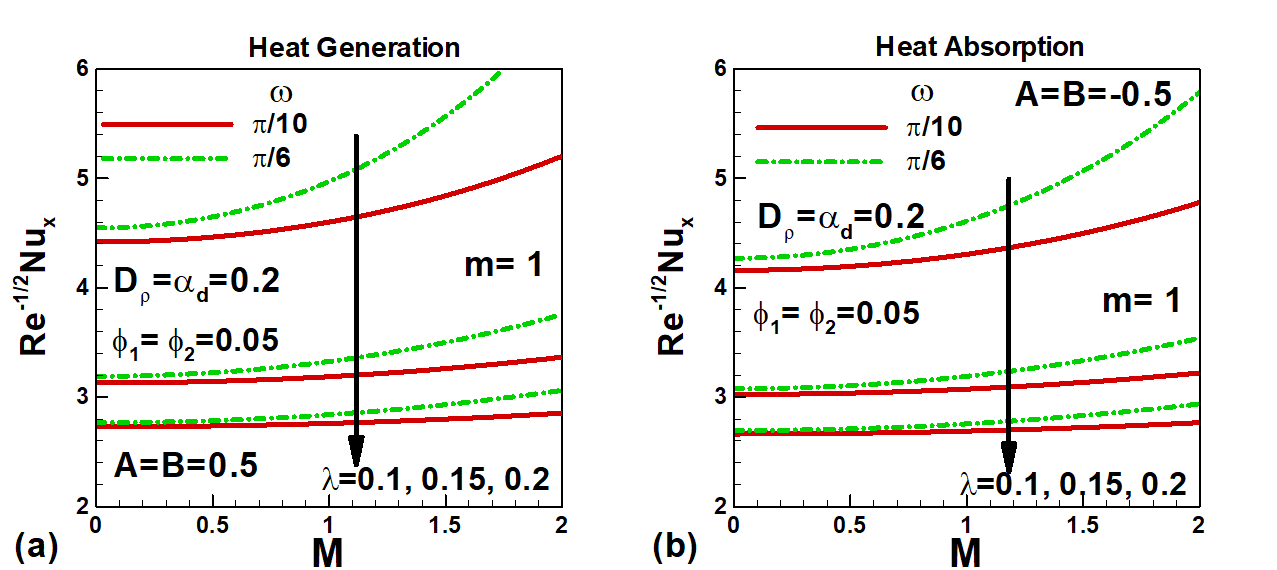}
\caption{Variation in Nusselt number with an inclined magnetic field for hybrid nanofluid near plane stagnation point on vertical stretching flat plate with heat (a) generation and (b) absorption}
\label{f-17}
\end{figure}

\begin{figure}[h]
\centering
\includegraphics[width=13cm, height=5.2cm]{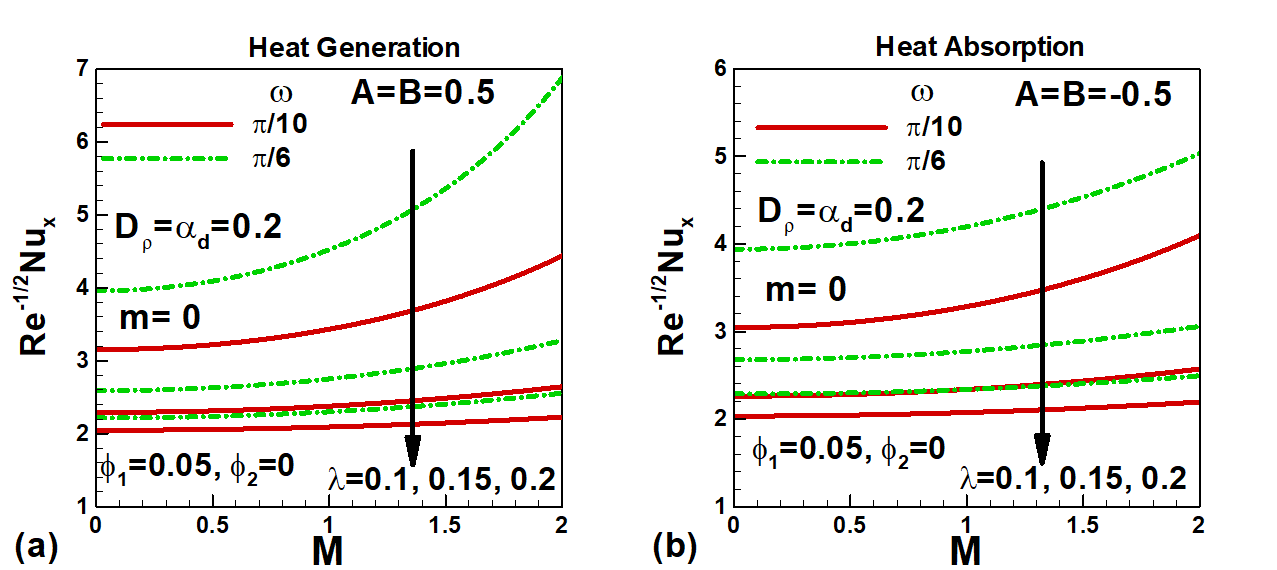}
\caption{Variation in Nusselt number with an inclined magnetic field for nanofluid on a horizontally stretching flat plate subjected to both (a) heat generation and (b) absorption}
\label{f-18}
\end{figure}

\begin{figure}[h]
\centering
\includegraphics[width=13cm, height=5.2cm]{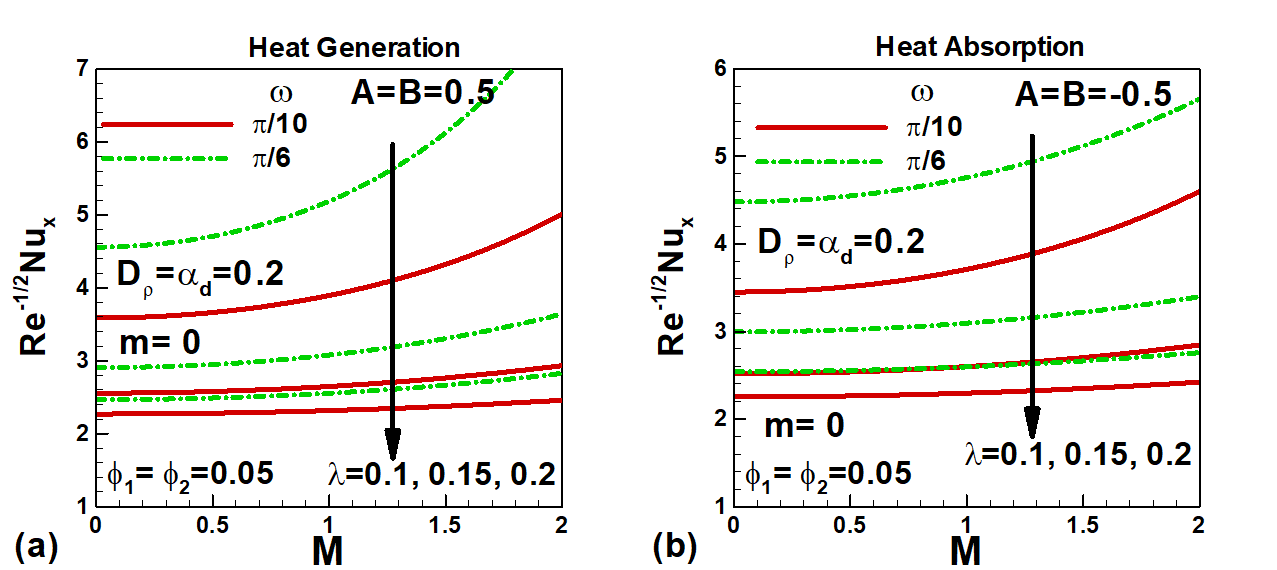}
\caption{Variation in Nusselt number with an inclined magnetic field for hybrid nanofluid on a horizontally stretching flat plate subjected to both (a) heat generation and (b) absorption}
\label{f-19}
\end{figure}

The variations in skin friction contrary to the magnetic field variable for both a nanofluid and a hybrid nanofluid are elucidated in figures (\ref{f-14}) and (\ref{f-15}). It's observed that a rise in $\lambda$ leads to stronger coupling in the fluid and the surface, typically enhancing the skin friction coefficient $Re^2 C_{f}$. An explanation for this occurrence is that a higher fluid velocity leads to a more significant impact of viscous forces at the boundary layer, resulting in an increased skin friction coefficient. Furthermore, the coefficient exhibits an increase with $M$ during both heat generation and absorption in both figures, indicating a higher resistance at the wedge's surface due to magnetic effects. The inclined magnetic field introduces additional forces, likely the Lorentz force, which can act as an opposing force opposing the fluid movement. Consequently, the skin friction coefficient rises for magnetic field strength increases, demonstrating the substantial influence of magnetic effects on the flow-surface interaction.
The figures (\ref{f-16}) and (\ref{f-17}) illustrate the behavior of the local Nusselt number in response to 
deviations in the velocity ratio parameter $\lambda$ for nanofluids and hybrid nanofluids. In both cases, the figures reveal that as $\lambda$ rises, the local Nusselt number $(Re^{-1/2} Nu_{x})$ drops for both heat generation and absorption cases. This trend suggests that a higher velocity ratio diminishes the relative effectiveness of convective cooling or heating at the surface. This behavior can be attributed to the configuration of a thinner thermal boundary layer as the fluid velocity increases, which results in reduced heat exchange between the fluid and the surface. The similar patterns observed in both figures indicate that this behavior is consistent regardless of the nanofluid composition, emphasizing the universal impact of the velocity ratio on the local Nusselt number in the context of nanofluid and hybrid nanofluid flows over wedge surfaces.
In Figures (\ref{f-18}) and (\ref{f-19}), the analysis of the local Nusselt number $(Re^{-1/2} Nu_{x})$ for both nanofluids and hybrid nanofluids influenced by an inclined magnetic field on a horizontally stretching flat plate is presented. In both figures, the results demonstrate that escalation in the velocity ratio parameter $\lambda$ induces a dwindling in the local Nusselt number $(Re^{-1/2} Nu_{x})$ during both heat generation and absorption scenarios. This observation suggests that the convective heat transfer efficiency diminishes as the fluid velocity increases relative to the stretching surface. The effect is consistent for both the simple nanofluid and the more complex hybrid nanofluid, indicating that the inclusion of $SiO_2$ nanoparticles does not alter the fundamental attitude of the fluid under the magnetic field and heat transfer conditions studied. The universal decrease in Nusselt number with increasing velocity ratio highlights the importance of considering the impact of fluid velocity on heat transfer efficiency in these complex flow scenarios.

\section{Outcomes}
In this comprehensive study on dusty hybrid nanofluids, several key findings have emerged:
\begin{enumerate}
    \item Increasing the magnetic field intensity $(M)$ consistently brings decline in the dimensionless velocity of both nanofluids and hybrid nanofluids near the stagnation point.
    \item Magnetic fields emit high levels of energy and enhance the thickness of thermal boundary layers.
    \item The sensitivity of the skin friction coefficient changes the velocity ratio parameter $(\lambda)$.
    \item An increase in $\lambda$ correlates with a rise in the drag coefficient, indicating a intensified connection between the fluid flow and the surface of the wedge.
    \item Thermal convection efficiency diminishes as the fluid velocity increases relative to the stretching surface.
\end{enumerate}
These outcomes collectively contribute to a deeper understanding of the complex interplay between magnetic fields, fluid dynamics, and thermal behavior in dusty hybrid nanofluid systems, offering valuable insights for researchers and engineers in various applications.

\bibliography{mybibfile}
\end{document}